\documentclass[12pt,twoside]{article}
\usepackage{epic}
\usepackage{eepic}
\usepackage[mathscr]{eucal}
\usepackage{amsmath,amsfonts,amssymb,amsthm}
\usepackage[matrix,arrow]{xy}
\usepackage{times}
\usepackage{graphicx}
\usepackage{epstopdf}
\usepackage{dcolumn}
\usepackage{hyperref}

\def\d_Vphi{\text{d}_V\hspace{-0.06em}\phi}
\def\d_Vphibar{\text{d}_V\hspace{-0.06em}\bar\phi}
\def\d_Vxi{\text{d}_V\hspace{-0.06em}\xi}

\def\ndelta{\delta\hspace{-0.50em}\slash\hspace{-0.05em} }

\def\be{\begin{eqnarray}}
\def\ee{\end{eqnarray}}
\def\beann{\begin{eqnarray*}}
\def\eeann{\end{eqnarray*}}
\def\beq{\begin{equation}}
\def\eeq{\end{equation}}
\def\ba{\begin{array}}
\def\ea{\end{array}}
\def\ben{\begin{enumerate}}
\def\een{\end{enumerate}}
\def\bea{\begin{eqnarray}}
\def\eea{\end{eqnarray}}

\def\5{\bar }
\def\6{\partial }
\def\7{\hat }
\def\4{\tilde }
\advance\textheight\topskip
\renewcommand{\tilde}{\widetilde}
\renewcommand{\hat}{\widehat}

\newcommand{\bref}[1]{\textbf{\ref{#1}}}

\newcommand{\dd}{\partial}
\renewcommand{\d}{\partial}

\renewcommand{\leq}{\,{\leqslant}\,}

\newcommand{\binner}[2]{%
  {\langle}\kern-4.15pt{\langle}#1{,}\,#2{\rangle}\kern-4.15pt{\rangle}}

\newcommand{\half}{\mathchoice{%
    \ffrac{1}{2}}{\frac{1}{2}}{\frac{1}{2}}{\frac{1}{2}}}

\newcommand{\ffrac}[2]{\raisebox{.5pt}%
  {\footnotesize$\displaystyle\frac{#1}{#2}$}\kern1pt}

\newcommand{\dover}[2]{\ffrac{\dd #1}{\dd #2}}

\newcommand{\ddl}[2]{\ffrac{\dd #1}{\dd #2}}

\def\cF{\mathcal{F}}

\def\cL{\mathcal{L}}

\def\cQ{\mathcal{Q}}

\def\cX{\mathcal{X}}
\def\cY{\mathcal{Y}}

\DeclareFontFamily{OT1}{rsfs}{} \DeclareFontShape{OT1}{rsfs}{m}{n}{
<-7> rsfs5 <7-10> rsfs7 <10-> rsfs10}{}
\DeclareMathAlphabet{\mycal}{OT1}{rsfs}{m}{n}
\def\scri{{\mycal I}}%
\numberwithin{equation}{section} \makeatletter
\@addtoreset{equation}{section}

\begin{document}

\def\mytitle{BMS charge algebra}

\pagestyle{myheadings} \markboth{\textsc{\small Barnich, Troessaert}}{%
  \textsc{\small BMS charge algebra}} \addtolength{\headsep}{4pt}

\begin{flushright}\small
ULB-TH/11-10\end{flushright}

\begin{centering}

  \vspace{1cm}

  \textbf{\Large{\mytitle}}



  \vspace{1.5cm}

  {\large Glenn Barnich$^{a}$ and C\'edric Troessaert$^{b}$}

\vspace{.5cm}

\begin{minipage}{.9\textwidth}\small \it \begin{center}
   Physique Th\'eorique et Math\'ematique\\ Universit\'e Libre de
   Bruxelles\\ and \\ International Solvay Institutes \\ Campus
   Plaine C.P. 231, B-1050 Bruxelles, Belgium \end{center}
\end{minipage}

\end{centering}

\vspace{1cm}

\begin{center}
  \begin{minipage}{.9\textwidth}
    \textsc{Abstract}. The surface charges associated with the
    symmetries of asymptotically flat four dimensional spacetimes at
    null infinity are constructed. They realize the symmetry algebra
    in general only up to a field-dependent central extension that
    satisfies a suitably generalized cocycle condition. This extension
    vanishes when using the globally well defined BMS algebra. For the
    Kerr black hole and the enlarged BMS algebra with both
    supertranslations and superrotations, some of the
    supertranslations charges diverge whereas there are no divergences
    for the superrotation charges. The central extension is
    proportional to the rotation parameter and involves divergent
    integrals on the sphere.
  \end{minipage}
\end{center}

\vfill

\noindent
\mbox{}
\raisebox{-3\baselineskip}{%
  \parbox{\textwidth}{\mbox{}\hrulefill\\[-4pt]}}
{\scriptsize$^a$Research Director of the Fund for Scientific
  Research-FNRS. E-mail: gbarnich@ulb.ac.be\\$^b$ Research Fellow of
  the Fund for Scientific Research-FNRS. E-mail: ctroessa@ulb.ac.be}

\thispagestyle{empty}
\newpage

\begin{small}
{\addtolength{\parskip}{-1.5pt}
 \tableofcontents}
\end{small}
\newpage

\section{Introduction}
\label{sec:introduction}

In the study of gravitational radiation in the early sixties
\cite{Bondi:1962px,Sachs:1962wk}, it turned out that the asymptotic
symmetry group at null infinity in four dimensions is not the
Poincar\'e group, but an enhanced group where translations are
replaced by supertranslations. We have recently shown
\cite{Barnich:2009se,Barnich:2010eb} that on the level of the algebra,
one can consistently allow for infinitesimal superrotations as well
and have worked out the transformation laws of the functions
parametrizing solution space. The resulting symmetry algebra
$\mathfrak{bms}_4$ is an extension of the Poincar\'e algebra that
contains two copies of the Virasoro algebra. It thus follows that
asymptotically flat general relativity in four dimensions is dual to
an extended conformal field theory. 

An important element that is missing in our analysis is the
construction of surface charges associated to $\mathfrak{bms}_4$
together with their transformation laws. This is a notoriously
difficult task as the surface charges are non-conserved and
non-integrable at null infinity \cite{Wald:1999wa}. It is the purpose
of the present paper to fill this gap.  What we are especially
interested in are the transformation properties of the surface
charges. Indeed, in the anti-de Sitter case in three dimensions, the
central extension \cite{Brown:1986nw} that appears has been used to
argue for a microscopic explanation of the Bekenstein-Hawking entropy
of the BTZ black hole \cite{Strominger:1998eq}. A similar analysis has
been applied in the near-horizon limit of an extreme four dimensional
Kerr black hole \cite{Guica:2008mu,Bredberg:2011hp}.

The main result of our paper is the construction of the field
dependent central extension that generically occurs in the charge
algebra at null infinity.

When the symmetry algebra is the standard, globally well-defined BMS
algebra, we show that the extension vanishes. When using the
extended BMS algebra with both supertranslations and superrotations
instead and evaluating for a Kerr black hole, some of the
supertranslation charges as well as the non-vanishing extension
involves divergent integrals on the $2$-sphere.

Whether our results can be used in the context of a microscopic
derivation of the entropy of a Kerr black hole thus depends on the
question of how to regularize the divergent integrals that occur and
how to extract meaningful answers. Some comments on this problem are
provided at the end of the paper. 

A more complete and general theory for surface charges in the
non-integrable case, together with a better understanding of how they
generate the asymptotic symmetry transformations in a Dirac or Peierls
bracket, is also needed. We hope to address some of these issues
elsewhere.

\section{Summary of previous results}
\label{sec:mathfr-surf-charg}

\subsection{General expressions for surface charge one-forms from
  linearized theory}
\label{sec:linearized-theory}
~
Our starting point is the covariant approach to surface charges and
their algebra developed in \cite{Barnich:2001jy} (see also
\cite{Barnich:2003xg,Barnich:2007bf}).  In particular, for pure
Einstein gravity with or without a cosmological constant, it has been
shown in \cite{Barnich:2004ts} that for the linearized theory,
described by $h_{\mu\nu}$ around a background $g_{\mu\nu}$, the
conserved surface charges are completely classified by the Killing
vectors $\xi^\mu$ of the metric $g_{\mu\nu}$.  These charges only
depend on the Einstein equations of motion and not on the choice of
Lagrangian. They form a representation of the Lie algebra of Killing
vectors of $g_{\mu\nu}$. Their explicit expression coincides with
formulas derived earlier in \cite{Abbott:1981ff} and is given by
\begin{multline}
  \label{eq:29}
 \ndelta \cQ_\xi[h, g]= \frac{1}{16
    \pi G}\int_{S}\,(d^{n-2}x)_{\mu\nu}\, \sqrt{- g}\Big[\xi^\nu D^\mu h
  -\xi^\nu D_\sigma h^{\mu\sigma} +\xi_\sigma D^\nu
  h^{\mu\sigma}
  \\
  +\frac{1}{2}h D^\nu\xi^\mu +\half
  h^{\nu\sigma}(D^\mu\xi_\sigma-D_\sigma\xi^\mu)
  -(\mu\leftrightarrow \nu)\Big]\,,
\end{multline}
where
\[(d^{n-k}x)_{\nu\mu}=\frac{1}{k!(n-k)!}
\epsilon_{\nu\mu\alpha_1\dots\alpha_{n-2}}
dx^{\alpha_1}\wedge\dots\wedge dx^{\alpha_{n-2}},\quad \epsilon_{01\dots
  n-1}=1.\]

In view of these universal properties of the surface charges in the
linearized theory, we use them in the context of asymptotically flat
four dimensional spacetimes at null infinity. Whereas there is no
issue with integrability in the linearized theory, in the full
interacting theory with prescribed asymptotics, the expressions are
one-forms on solution space indexed by asymptotic symmetries and one
has to face the question whether these one-forms are integrable, i.e.,
whether one can construct suitable ``Hamiltonians'' for
them~\cite{Wald:1999wa}. This explains the notation $\ndelta$ in
\eqref{eq:29}.

More precisely, in the case at hand, $S$ is a spherical cross-section
of future or past null infinity, ``Scri'' denoted by $\scri$.  The
metric $g_{\mu\nu}$ is an asymptotically flat solution to Einstein's
equations, $h_{\mu\nu}$ a solution to the linearized equations at
$g_{\mu\nu}$ and $\xi^\mu$ a space-time vector realizing the
$\mathfrak{bms}_4$ algebra on asymptotically flat
spacetimes. Throughout, we will use the conventions of
\cite{Barnich:2010eb} to which we refer for further details. We thus
have $n=4$, the coordinates are $u,r$ and $x^A=\theta,\phi$, with $S$
the $2$-sphere at $u=u_0$ and $r=cst\to\infty$, i.e., the limits of
integration are $0\leq \theta\leq \pi$ and $0\leq \phi\leq 2\pi$. We
will also use the notation $\int d^2\Omega^\varphi=\int dx^2
dx^3\sqrt{\bar\gamma}=\int^{2\pi}_0d\phi\int_0^\pi d\theta\sin\theta\, 
e^{2\varphi}$ below.

\subsection{Solution space}
\label{sec:solution-space}
~
Asymptotically flat metrics solving Einstein's
equation are of the form
\begin{equation}
  \label{eq:2}
  ds^2=e^{2\beta}\frac{V}{r} du^2-2e^{2\beta}dudr+
g_{AB}(dx^A-U^Adu)(dx^B-U^Bdu)\,,
\end{equation}
where
\begin{equation}
  \label{eq:96}
  g_{AB}=r^2\bar\gamma_{AB}+rC_{AB}+D_{AB}+\frac{1}{4} \bar\gamma_{AB}
C^C_DC^D_C+o(r^{-\epsilon})\,.
\end{equation}
The background metric is 
\begin{equation}
\begin{split}
  \bar\gamma_{AB}dx^Adx^B=e^{2\varphi}(d\theta^2+\sin\theta
  d\phi^2)=e^{2\tilde\varphi}d\zeta d\bar \zeta\label{eq:32},\\
  \zeta=\cot{\frac{\theta}{2}}e^{i\phi},\quad
  \tilde\varphi=\varphi-\varphi_0,\quad \varphi_0=\ln P, \quad
  P=\half(1+\zeta\bar\zeta)\,.
\end{split}
\end{equation}
We assume for
simplicity that $\varphi,\tilde\varphi$ do not depend on $u$,
$\varphi=\varphi(x^A)$. Indices on $C_{AB},D_{AB}$ are raised with the
inverse of $\bar\gamma_{AB}$ and $C^A_A=0=D^A_A$. In addition $\d_u
D_{AB}=0$ and the news tensor is $N_{AB}=\d_u C_{AB}$. Furthermore,
\begin{equation}
  \label{eq:4}
  \beta=-\frac{1}{32}r^{-2}C^A_BC^B_A-\frac{1}{12}
    r^{-3} C^A_BD^B_A+o(r^{-3-\epsilon})\,,
\end{equation}
\begin{multline}
  \label{eq:62}
  g_{uA}=\frac{1}{2}\bar D_BC^{B}_{A}+
\frac{2}{3}r^{-1}\Big[(\ln r+
  \frac{1}{3})\bar
D_BD^{B}_{A}\\+\frac{1}{4} C_{AB} \bar D_CC^{CB}
  +N_A\Big]
  +o(r^{-1-\varepsilon})\,,
\end{multline}
where $\bar D_A$ is the covariant derivative associated to
$\bar\gamma_{AB}$ and $N_A(u,x^A)$ is
the angular momentum aspect;
\begin{equation}
  \label{eq:69}
  \frac{V}{r}  =-\frac{1}{2} \bar
  R + r^{-1}2M + o(r^{-1-\epsilon}),
\end{equation}
where $\bar R$ is the scalar curvature of $\bar D_A$, $\bar
R=2e^{-2\varphi}-2\bar\Delta\varphi$ with $\bar \Delta$ the Laplacian
for $\bar\gamma_{AB}$ and $M(u,x^A)$ is the mass aspect.  Finally, the
evolution of the mass and angular momentum aspects in retarded time
$u$ is determined by
\begin{equation}
  \label{eq:3}
  \d_uM=-\frac{1}{8} N^A_BN^B_A
 +\frac{1}{8}
  \bar \Delta \bar R 
  +\frac{1}{4}\bar D_A\bar D_C N^{CA},
\end{equation}
\begin{multline}
  \label{eq:78ter}
\d_uN_A =\d_AM+\frac{1}{4}C_A^B\d_B\bar
  R +\frac{1}{16}\d_A\big[N^B_C C^C_B\big]
-\frac{1}{4} \bar D_AC^C_BN^B_C\\
-\frac{1}{4}\bar D_B\big[C^{B}_{C}N^C_{A}-N^B_CC^C_A\big]-\frac{1}{4}
  \bar D_B \big[ \bar D^B \bar D_CC^C_A -\bar D_A \bar
  D_CC^{BC}\big].
\end{multline}

To summarize, coordinates on solution space to the order we need, are
given by
\begin{equation}
\cX^\Gamma\equiv \{C_{AB},N_{AB},D_{AB},M,N_A\}\label{eq:6}.
\end{equation}
Using the evolution equation equations \eqref{eq:3},\eqref{eq:78ter},
the definition of the news and the $u$ independence of $D_{AB}$, all
these fields can be taken at fixed $u=u_0$ and thus depend only on
$x^A$, except for the news which contains an arbitrary $u$ dependence, 
$N_{AB}=N_{AB}(u,x^A)$.

We consider $\varphi$ to be part of the gauge fixing which we do
not vary at this stage. It thus follows that $h_{\mu\nu}$ is entirely
determined to the order we need by $\delta \cX^\Gamma$.

Note in particular that \eqref{eq:3} controls the mass loss as shown
in \cite{Bondi:1962px,Sachs:1962wk}. By integrating over the sphere,
one finds $\d_u\int_S d^2\Omega\, M=-\frac{1}{8}\int d^2\Omega\,
N^A_BN^B_A$. By definition, the left hand side is the Bondi mass
whereas, in spherical or in stereographic coordinates, the right hand
side can easily be seen to be negative and zero if and only if the
news tensor vanishes. It follows that the Bondi mass is constant
unless the news tensor is non-vanishing in which case the Bondi mass
can only decrease in retarded time $u$ .

\subsection{Asymptotic symmetry algebra and its action on solution
  space}
\label{sec:asympt-symm-algebra}

Let $s=(T,Y)\in \mathfrak{bms}_4$ denote a generic element of the
symmetry algebra, which consists of the semi-direct sum of the Lie
algebra $Y^A\d_A$ of conformal Killing vectors of the $2$ sphere,
``infinitesimal superrotations'', acting in a suitable way on
infinitesimal supertranslations which are parametrized by arbitrary
functions $T=T(x^A)$, $[s_1,s_2]\equiv
[(T_1,Y_1),(T_2,Y_2)]=(\hat T,\hat Y)$, with
\begin{equation}
  \label{eq:9}
  \hat Y=Y_1^B\d_B Y_2^A- (1\leftrightarrow 2) \,,\quad 
  \hat T=Y_1^A \d_A T_2 -\half  \bar D_A Y_1^A T_2 -(1\leftrightarrow 2). 
\end{equation}
In stereographic coordinates $\zeta,\bar \zeta$, the algebra may be
realized through the vector fields $y=Y(\zeta)\d $, $\bar y=\bar
Y(\bar \zeta)\bar \d$, with $\d= \dover{}{\zeta}$, $\bar
\d=\dover{}{\bar \zeta}$. Let $T(\zeta,\bar\zeta)=\tilde
T(\zeta,\bar\zeta)e^{\tilde\varphi}$. In the language used in the
study of the Virasoro algebra (see e.g.~\cite{Fuks:1986}), the
conformal Killing vectors act on tensor densities $\cF_{\half,\half}$
of degree $(\half,\half)$, $t=\tilde T(\zeta,\bar\zeta)e^{\tilde
  \varphi}(d\zeta)^{-\half}(d\bar\zeta)^{-\half}$ through
\begin{eqnarray}
\rho(y) t =(Y\d \tilde T-\half \d Y
\tilde T)e^{\tilde \varphi}(d\zeta)^{-\half}(d\bar\zeta)^{-\half}\,,\\
\rho(\bar y) t=(\bar
Y\bar \d \tilde T-\half \bar \d \bar Y
\tilde T)e^{\tilde \varphi}(d\zeta)^{-\half}(d\bar\zeta)^{-\half}\,. \label{eq:9a}
\end{eqnarray}
The algebra $\mathfrak{bms}_4$ is then the semi-direct sum of the
algebra of vector fields $y,\bar y$ with the abelian ideal
$\cF_{\half,\half}$, the bracket being induced by the module action,
$[y,t]=\rho(y)t$, $[\bar y,t]=\rho(\bar y) t$.
When expanding $y=a^nl_n$, $\bar y=\bar a^n\bar l_n$, $t=b^{m,n} T_{m,n}$, 
where
\begin{equation}
  \label{eq:10}
  l_n=-\zeta^{n+1}\d,\quad \bar l_n=-\bar \zeta^{n+1}\bar\d,\quad 
T_{m,n}=\zeta^{m}\bar\zeta^{n}e^{\tilde \varphi}
(d\zeta)^{-\half}(d\bar\zeta)^{-\half}\,,
\end{equation}
with $m,n\dots\in \mathbb Z$, 
the enhanced symmetry algebra reads
\begin{eqnarray}
  \label{eq:37}
  [l_m,l_n]=(m-n)l_{m+n},\quad [\bar l_m,\bar l_n]=(m-n)\bar
  l_{m+n},\quad [l_m,\bar l_n]=0, \cr
 [l_l,T_{m,n}]=(\frac{l+1}{2}-m)T_{m+l,n},
\, [\bar l_l,T_{m,n}]= (\frac{l+1}{2}-n)T_{m,n+l},\, [T_{m,n},T_{o,p}]=0.
\end{eqnarray}
The Poincar\'e algebra is the
subalgebra spanned by the generators $T_{0,0}$, $T_{0,1}$, $T_{1,0}$,
$T_{1,1}$ for ordinary translations and $l_{-1},l_0,l_1,\bar
l_{-1},\bar l_{0},\bar l_1$ for ordinary (Lorentz) rotations.

The space-time vectors $\xi=\xi[s;g]$ that realize the asymptotic
symmetry algebra $\mathfrak{bms}_4$ in the modified bracket,
\begin{multline}
[\xi[s_1;g],\xi[s_2;g]]_M\equiv[[\xi[s_1;g],\xi[s_2;g]]-\delta^g_{\xi[s_1;g]}\xi[s_2;g]+
\delta^g_{\xi[s_2;g]}\xi[s_1;g]=\\=\xi[[s_1,s_2];g]\,,\label{eq:8}
\end{multline}
with $\delta^g_{\xi}g_{\mu\nu}=\cL_\xi g_{\mu\nu}$, are explicitly
given by
\begin{gather}
  \label{eq:26}
\left\{\begin{array}{l}
  \xi^u=f,\\
\xi^A=Y^A+I^A, \quad  I^A=- f_{,B} \int_r^\infty dr^\prime(
e^{2\beta} g^{AB}),\\
\xi^r=-\half r (\bar D_A\xi^A-f_{,B}U^B),
\end{array}\right.
\end{gather}
where $Y^A=Y^A(x^B)$ are conformal Killing vectors of the $2$ sphere,
$f=e^\varphi T+\half u\psi$ with $ \psi= \bar D_A Y^A$. 

Their action on solution space can be worked out to be
\begin{equation}
  \label{eq:16}
  -\delta_s C_{AB}=[f\d_u+\cL_Y -\half\psi] C_{AB}-2\bar D_A\bar D_B f
+\bar \Delta f\bar\gamma_{AB}\,.
\end{equation}
\begin{equation}
  \label{eq:22}
  -\delta_s N_{AB}= [f\d_u + \cL_Y] N_{AB}-(\bar D_A\bar D_B
   \psi-\half \bar \Delta \psi \bar\gamma_{AB})\,,
\end{equation}
\begin{equation}
  \label{eq:19}
  -\delta_s D_{AB}=\cL_Y D_{AB}\,, 
\end{equation}
\begin{multline}
  \label{eq:34}
  -\delta_s M =[f\d_u +Y^A\d_A+\frac{3}{2}
  \psi]M\\+\frac{1}{4}\d_u[ \bar D_C \bar D_B f C^{CB} +2\bar
  D_B f\bar D_C C^{CB}] -\frac{1}{4}\bar D_A \psi\bar D_B C^{BA}
  +\frac{1}{4}\d_A f\d^A\bar R\,,
\end{multline}
\begin{multline}
  \label{eq:33}
  -\delta_s N_A=[f\d_u+\cL_Y+\psi]N_A -\half [ 
  \bar D_B  \psi +\psi
  \bar  D_B ] D^B_A\\
  +3\bar D_A f M -\frac{3}{16}\bar D_A f
  N^B_CC^C_B
+\half \bar D_B f N^B_C C^C_A
-\frac{1}{32}\bar D_A\psi
  (C^B_CC^C_B)\\+ \frac{1}{4} (\bar D_Bf \bar R+\bar D_B \bar\Delta
  f)C^B_A -\frac{3}{4}\bar D_B f(\bar D^B\bar D_C C^C_A-\bar
  D_A\bar D_C C^{BC})
  \\+\half (\bar D_A\bar D_B f
-\frac{1}{2}\bar \Delta f\bar\gamma_{AB} )\bar D_C C^{CB}+\frac{3}{8} \bar
  D_A(\bar D_C\bar D_B f C^{CB}) . 
\end{multline}

\subsection{Globally well-defined symmetry algebra}
\label{sec:glob-well-defin}

In the standard approach to the BMS symmetry algebra in general
relativity, one restricts oneself to globally well-defined
transformations on the sphere. This amounts to considering only
$l_m,\bar l_n$, with $m,n$ taking the values $-1,0,1$. At the same
time, the supertranslations are restricted to those that can be
expanded into spherical harmonics $Y_{lm}$. The supertranslation
generators are then, $t=c^{lm} \cY_{lm}$ where
$\cY_{lm}=Y_{lm}(\zeta,\bar\zeta)(d\zeta)^{-\half}(d\zeta)^{-\half}$. The
commutation relations $[l_n,\cY_{lm}]$ have been worked out already in
\cite{Sachs2}. More general considerations on the transformation
properties of (spin weighted) spherical harmonics under Lorentz
transformations can be found in
\cite{goldberg:2155,held:3145,Penrose:1984}. For later use, let us denote the
standard, globally well-defined BMS algebra on the sphere by
$\mathfrak{bms}_4^{\rm glob}$. 

\section{Charge algebra}

\subsection{Charges for asymptotically flat spacetimes at null
  infinity}
\label{sec:surface-charge-one}

Using the data summarized in the previous section and inserting into
\eqref{eq:29} gives, after a lengthy computation whose main steps are
summarized in the appendix,
\begin{equation}
  \label{eq:36}
   \ndelta\cQ_\xi[\delta\cX,\cX] = \delta \left(Q_{s}[\cX]\right)+
   \Theta_{s}[\delta\cX,\cX]\, ,
 \end{equation}
 where the integrable part of the surface charge one-form is
 given by 
 \begin{equation}
   \label{eq:7}
   Q_{s}[\cX]=\frac{1}{16 \pi G}\int
   d^2 \Omega^\varphi\,\Big[ 4 f M+ Y^A \big(2 N_A + 
\frac{1}{16}\partial_A (C^{CB} C_{CB})\big) \Big]\,,
 \end{equation}
and the non-integrable part is due to the news tensor,
\begin{equation}
  \Theta_{s}[\delta\cX,\cX]=\frac{1}{16 \pi G}\int
  d^2 \Omega^\varphi\,
  \Big[ \frac{f}{2} N_{AB} \delta C^{AB}\Big]\,.
\end{equation}

The separation into an integrable and non-integrable part in
\eqref{eq:36} is not uniquely defined as this equation also holds in
terms $Q^\prime_s=Q_s-N_s$, $\Theta^\prime_s=\Theta_s+\delta N_s$ for
some $N_s[\cX]$. 

These charges are very similar and should be compared to those
proposed earlier in \cite{Wald:1999wa} in the context of a closely
related, but slightly different approach to asymptotically flat
spacetimes.

\subsection{Charges as representations of the symmetry
  algebra}

In the integrable Hamiltonian case
\cite{Regge:1974zd,Brown:1986ed,Brown:1986nw}, it has been shown that
the asymptotic symmetry algebra is represented through the Dirac
bracket of the surface charges, up to a central extension,
\begin{equation}
\{Q^H_{s_1},Q^H_{s_2}\}^*=-\delta_{s_2} Q^H_{s_1}=Q^H_{[s_1,s_2]}+
K^H_{s_1,s_2}\label{eq:10b}
\end{equation}
where $K^H_{s_1,s_2}$ is a Lie algebra 2-cocycle (with values in the
real numbers). In the covariant approach, one can show a similar
result \cite{Barnich:2001jy,Barnich:2007bf} . More precisely, when the
charges are integrable, one can show that $-\delta_{s_2}
Q_{s_1}=Q_{[s_1,s_2]}+ K_{s_1,s_2}$ where $K_{s_1,s_2}$ is again a Lie
algebra $2$-cocycle taking values in the real numbers. When using the
equivalence of the Hamiltonian and the covariant approaches, one can
infer that this coincides with the Dirac bracket
$\{Q_{s_1},Q_{s_2}\}^*$ of the charges.

In the non integrable case, we propose as a definition
\begin{equation}
\left\{Q_{s_1},Q_{s_2}  \right\}^*[\cX] = (-\delta_{s_2}) Q_{s_1}[\cX] +
\Theta_{s_2} [-\delta_{s_1} \cX,\cX]\,. \label{fund1}
\end{equation}
Whether this definition generically makes sense and defines a Dirac
bracket will be addressed elsewhere. The point we want to make is
that, in the case at hand, the right hand side can be shown to be
given by the charges for the commutators of the symmetries, up to a
field dependent central extension. Indeed, we will show in the
appendix that
\begin{equation}
\left\{Q_{s_1},Q_{s_2}  \right\}^*= Q_{[s_1,s_2]} +
K_{s_1,s_2}, \label{skew}
\end{equation}
where the field dependent central extension is 
\begin{multline}
K_{s_1,s_2}[\cX]= \frac{1}{32 \pi G}\int
   d^2 \Omega^\varphi\,
\Big[  ( f_1
  \d_A f_2 - f_2 \d_A f_1)\d^A \bar R+\\+C^{BC}  (f_1 \bar D_B \bar
  D_C \psi_2 - f_2 \bar D_B \bar D_C
  \psi_1) \Big]\,. \label{centralcha}
\end{multline}
This central extension satisfies the suitably generalized 
cocycle condition
\begin{equation}
\label{cocycle}
K_{[s_1,s_2],s_3} - \delta_{s_3} K_{s_1,s_2} + {\rm cyclic}\ (1,2,3)
=0.
\end{equation}
In fact, \eqref{skew} and \eqref{cocycle} imply the Jacobi identity
for the proposed bracket when the algebra element associated to
$K_{s_1,s_2}$ is central and thus generates no transformation. More
precisely, $\{\cdot,\cdot\}^*$ defines a Lie bracket for the elements
$Q_{s_1},K_{s_2,s_3}$ if one defines in addition that
$\{K_{s_1,s_2},Q_{s_3}\}^* = -\delta_{s_3}
K_{s_1,s_2}=-\{Q_{s_3},K_{s_1,s_2}\}^*$ and $\{K_{s_1,s_2},
K_{s_3,s_4}\}^*=0$.

When defining as before,
$\left\{Q^\prime_{s_1},Q^\prime_{s_2} \right\}^*[\cX] =
(-\delta_{s_2}) Q^\prime_{s_1}[\cX] + \Theta^\prime_{s_2}
[-\delta_{s_1} \cX,\cX]$, one gets
$  \left\{Q^\prime_{s_1},Q^\prime_{s_2}  \right\}^*= Q^\prime_{[s_1,s_2]}+
K^\prime_{s_1,s_2}$,
where 
\begin{equation}
K^\prime_{s_1,s_2}=K_{s_1,s_2}+\delta_{s_2}
N_{s_1}-\delta_{s_1} N_{s_2}+N_{[s_1,s_2]}.\label{eq:48}
\end{equation}
Note that $\delta_{s_2} N_{s_1}-\delta_{s_1} N_{s_2}+N_{[s_1,s_2]}$ is
a trivial field dependent $2$-cocycle in the sense that it
automatically satisfies the cocyle condition \eqref{cocycle}.

{\bf Discussion:}
\begin{itemize}
\item The proved equality between the right hand sides of \eqref{fund1} and
\eqref{skew} controls the non-conservation of the charges. Indeed, by
taking $s_2=( T=1, Y^A=0)$ and $s_1=s$ we find from $\frac{d}{du}
Q_{s}= \ddl{}{u} Q_s-\delta_{1,0} Q_s$ that
\begin{multline}
  \label{eq:12b}
  \frac{d}{du} Q_{s}=-\frac{1}{32\pi G}\int d^2\Omega^\varphi\Big[
  N^{AB}\big([f\d_u +\cL_Y -\half \psi] C_{AB} -2 \bar D_A\bar D_B
  f\big)+\\+ \d_A f \d^A \bar R+ C^{BC} \bar D_B\bar D_C\psi\Big]\,.
\end{multline}
The standard result that the mass loss is positive and vanishes only
in the absence of news then follows by taking $s=(T=1,Y^A=0)$. 

\item It also follows that on the sphere, the standard $\mathfrak{bms}^{\rm
  glob}_4$ charges are all conserved in the absence of news.

\item In the case of the standard $\mathfrak{bms}_4^{\rm glob}$
  algebra on the sphere, there are no divergences provided the
  asymptotic solutions $\cX$ are well-defined. The central charge
  $K_{s_1,s_2}$ vanishes and the representation of the asymptotic
  symmetry algebra through the charges simplifies to
  \begin{equation}
    \label{eq:21}
    \left\{Q_{s_1},Q_{s_2}  \right\}^*= Q_{[s_1,s_2]} 
  \end{equation}
  To the best of our knowledge, even in this well-studied case this
  representation theorem is a new result that does so far not exist in
  any other formulation of the problem.
\end{itemize}

\section{Charges and central extension for the Kerr black hole}
\label{sec:centr-extens-kerr}

We will now take as a background metric the standard metric on the
sphere, i.e., $\varphi=0$. By following \cite{0264-9381-20-19-302} and
choosing the radial coordinate appropriately, one can put the Kerr
black hole in BMS coordinates. As shown in the appendix, the Kerr
solution $\cX^{Kerr}$ corresponds to $M(u,\theta,\phi)=M$, with $M$
the constant mass parameter of the Kerr black hole, $D_{AB}=0=N_{AB}$
while
\begin{eqnarray}
  \label{eq:5}
  C_{\theta\theta}=\frac{a}{\sin{\theta}},\quad
  C_{\phi\phi}=-a\sin{\theta},
\quad C_{\theta\phi}=0,\\
  N_\theta=3Ma\cos{\theta}+\frac{a^2}{8}\frac{\cos{\theta}}{\sin^3{\theta}},\quad
  N_\phi=-3aM\sin^2{\theta}\,.
\end{eqnarray}
Note that in the BMS gauge, $C_{\theta\theta}$ and $N_\theta$ are 
singular both on the north and the south pole. 

For the supertranslation charges, we find 
\begin{equation}
Q_{T_{m,n},0}[\cX^{Kerr}]=\frac{2M}{G}I_{m,n},\quad
I_{m,n}=\frac{1}{4\pi}\int d^2\Omega \frac{1}{1+\zeta\bar\zeta}
\zeta^{m}\bar\zeta^{n}\,. 
\end{equation}
A direct integration on the sphere gives $I_{m,n}=\delta^m_n
I(m)$, with 
\begin{equation}
I(m)=\frac{1}{4}\int^1_{-1}d\mu
\frac{(1+\mu)^{m}}{(1-\mu)^{m-1}}\label{eq:49}\,.
\end{equation}
We have $I(m)=I(1-m)$. In particular $I(0)=\half=I(1)$, so that the mass,
which is associated to the exact Killing vector $\d_u$ of the Kerr
solution and corresponds to
$T=1$, $Y=0$ and thus to $\half(T_{0,0}+T_{1,1})$, is given by
\begin{equation}
  \label{eq:18}
  Q_{T=1,Y=0}[\cX^{Kerr}]=\frac{M}{G}\,,
\end{equation}
as it should. For $m>1$ and $m<0$, the charges are not directly
well-defined as the integrals diverge. Note that in the case of the
globally well-defined BMS algebra, the supertranslations are expanded
in spherical harmonics, $T=c^{lm}Y_{lm}$. It follows that the only non
vanishing charge is \eqref{eq:18} while all other supertranslation
charges with $l>0$ vanish in this case.

The superrotations charges are given by 
\begin{equation}
  \label{eq:17}
  Q_{0,l_m}[\cX^{Kerr}]=-\delta^m_0\frac{iaM}{2G}.
\end{equation}
In particular, the standard angular momentum is associated to
the exact Killing vector $\d_\phi=-i(l_0-\bar l_0)$ of the Kerr
solution and is thus given by 
\begin{equation}
Q_{T=0,Y^\phi=1,Y^\theta=0}[\cX^{Kerr}]=-\frac{Ma}{G}\label{eq:13}\,,
\end{equation} 
as it should{\footnote{A discussion of the minus sign can for instance
    be found in \cite{Iyer:1994ys},
after equation (89).}

For the central extension, we find
\begin{equation}
  \label{eq:20}
  K_{(0,l_m),(0,l_n)}[\cX^{Kerr}]=0=K_{(0,\bar l_m),(0,\bar
    l_n)}[\cX^{Kerr}]=K_{(0,l_m),(0,\bar l_n)}[\cX^{Kerr}], 
\end{equation}
and 
\begin{eqnarray}
  \label{eq:27}
  K_{(0, l_l),(T_{m,n},0)}[\cX^{Kerr}]=\frac{a\,  l (l-1)(l+1)}{16 G} J_{m+l,n}\,,\\
K_{(0, \bar l_l),(T_{m,n},0)}[\cX^{Kerr}]=\frac{a\,  l (l-1)(l+1)}{16
  G} J_{m,n+l}\,,
\end{eqnarray}
with  
\begin{equation}
J_{m,n}=\frac{1}{4\pi} \int d^2\Omega
\frac{(1+\zeta\bar\zeta)^2}{\sqrt{\zeta^3\bar\zeta^3}} \zeta^{m}
\bar \zeta^{n}\label{eq:50}.
\end{equation}
The integration gives $J_{m,n}=\delta^m_n J(m)$ with 
\begin{equation}
J(m) =2\int_{-1}^1
d\mu\, \frac{(1+\mu)^{m-\frac{3}{2}}}{(1-\mu)^{m+\half}}, 
\end{equation}
and $J(m)=J(1-m)$. These integrals diverge for all integer values of
$m$.

\section{Discussion}
\label{sec:discussion}

The extended conformal field dual for four dimensional asymptotically
flat gravity is non-standard because the generator of time
translations is not related to the Virasoro generators $l_0$ and $\bar
l_0$ but to $\half (T_{0,0}+T_{1,1})$ instead. At the same time, the non trivial
central extension appears between the supertranslation and
superrotation generators, and not among the Virasoro generators alone.

To get to grips with these unusual features it is useful to review the
corresponding results for $\mathfrak{bms}_3$: a direct analysis of the
Dirac bracket algebra of the charges of asymptotically flat
space-times at null infinity in three dimensions
\cite{Barnich:2006avcorr} gives one non-centrally extended copy of the
Virasoro algebra with superrotation charges $L_m$ that act on the
commuting supertranslation charges $T_m$ with a (field independent)
central extension, $i[L_m,T_n]=(m-n) T_{m+n} + \frac{c^\prime}{12}
m(m^2-1)\delta^0_{m+n}$, where $c^\prime=\frac{3}{G}$ for the
Einstein-Hilbert action. In this case, there is no problem with
singularities since the boundary is a cylinder. Furthermore, the
relation to the asymptotically $AdS_3$ case sheds some light: starting
from two commuting copies $L^{\pm}_m$ of the Virasoro algebra with
central extensions $c^\pm$, the redefinition $L_m=L^+_m-L^-_{-m}$,
$T_m=\frac{1}{l}(L^+_m+L^{-}_{-m})$ implies that the $L_m$'s form a
copy of the Virasoro algebra with central charge $c^+-c^-$, the same
commutation relations between $L_m$ and $T_n$ as above with
$c^\prime=(1/l)(c^++c^-)$, while $i[T_m,T_n]=\frac{1}{l^2}( (m-n)
L_{m+n}+\frac{c^+-c^-}{12}m(m^2-1)\delta^0_{m+n})$. In the case of the
Einstein-Hilbert action where $c^\pm=\frac{3l}{2G}$, one then recovers
the $\mathfrak{bms}_3$ algebra in the limit $l\to \infty$ with zero
central extension for the Virasoro algebra of the $L_m$'s and the
above value $c^\prime=\frac{3}{G}$ between the superrotation and
supertranslation charges. From this point of view, the reason why the
central extensions for $\mathfrak{bms}_3$ in the pure gravity case
have this unusual structure is thus related to the fact that the
theory is a contraction of the standard conformal field theory of the
anti-Sitter case where left and right movers have the same central
charge.

A strategy to get a better understanding of the extended conformal
gravity dual in four dimensions is thus to first study the three
dimensional case in more detail. In particular, we will discuss
elsewhere the relation between the general asymptotically flat and
asymptotically anti-de Sitter solutions of three dimensional
gravity. The absence of black hole solutions in the purely
gravitational case with vanishing cosmological constant then forces
one to consider more exotic actions, such as the one for new massive
gravity \cite{Bergshoeff:2009hq} which admit asymptotically flat black
holes, to try to see what the analog of a Cardy formula has to look
like in order to reproduce the Bekenstein-Hawking entropy. One should
also directly study extended conformal field theories with
$\mathfrak{bms}_3$ symmetry by analysing its physically relevant
unitary irreducible representations. This has been partly done for the
$\mathfrak{gca}_2$ algebra \cite{Bagchi:2009my}, which is isomorphic
to the $\mathfrak{bms}_3$ algebra. Note however that the main
assumption that the energy should be bounded from below implies that
such a representation should have a lowest eigenvalue for $T_0$. We
plan to address some of these questions elsewhere.

In the same way than the modified Lie bracket needed to represent the
asymptotic symmetry algebra in the bulk space-time is the bracket of
the Lie algebroid naturally associated to gauge systems
\cite{Barnich:2010xq}, field dependent central extensions correspond
to Lie-algebroid 2-cocycles rather than to Lie algebra
2-cocycles. Note that, besides the standard central extensions in the
two Witt subalgebras, the $\mathfrak{bms}_4$ algebra does not admit
additional non trivial central extensions involving the
supertranslation generators, i.e., there are no additional non trivial
Lie algebra 2-cocycles with values in the real numbers (see
e.g.~\cite{barnich:2011ct}). This no-go result is circumvented here
because of the presence of the field $C_{AB}$.

The charges have been computed with respect to Minkowski space as a
background. In the context of the Kerr-CFT correspondence, it might be
more appropriate to choose another asymptotically flat solution as a
background, such as the extreme Kerr black hole for instance, or to
consistently restrict oneself to subclasses of solutions.

The proof that the charges represent the symmetry algebra up to a
field dependent central extension relies on the possibility to do
integrations by parts on the sphere.  This is of course problematic in
the case of divergent integrals. Then again, the central extension
seems interesting mainly in the case of a symmetry algebra consisting
of both supertranslations and superrotations where divergences are
unavoidable. 

A way to make sense of the divergent integrals could be to use the
theory of harmonic variables and distributions on the sphere
introduced in the context of harmonic superspace
\cite{Galperin:1985bj} (see also \cite{A.S.Galperin13674} for a
review) and applied to local conformal properties of the sphere in
\cite{saidi:1990xx,PhysRevD.46.777}. It would mean to probe solution
space through objects such as 
\begin{equation*}
\begin{gathered}
Q_{(T_{m,n},0)}[\cX](w^+,w^-)=\frac{1}{G}\int
dv\, M P^{-1} (\frac{w^-v^-}{w^-v^+})^m(\frac{w^+v^+}{w^+v^-})^n\,,\\
Q_{(0,l_m)}[\cX](w^-)=\frac{1}{G}\int dv\, (\frac{w^-v^-}{w^-v^+})^{m+1}
\Big[ \frac{u}{2}\d (P^{-1} M ) -\half N_\zeta-\frac{1}{64} \d
(C^{BC}C_{BC})\Big]\,.
\end{gathered}
\end{equation*}
The previous charges are then recovered for $w^-_1=0, w^-_2=1$ and
$w^+_1=1,w^+_2=0$.

An alternative to the approach sketched in the previous paragraph
consists in mapping the problem from the very beginning from the
standard to the Riemann sphere and use more standard conformal field
theory techniques. The formulas to do so are well known in the general
relativity literature (see e.g.~\cite{Geroch:1977aa,Ashtekar:1987tt})
since finite local conformal transformations of the two dimensional
part of metric remain as an ambiguity in Penrose's definition of
asymptotically flat spacetimes \cite{PhysRevLett.10.66}. In the
current set-up, the relevant formulas can be obtained by integrating
the infinitesimal transformation properties of the coordinates on
solution space under a local shift of the conformal factor
$-\delta\varphi=\omega$ worked out in \cite{Barnich:2010eb}.

\section*{Acknowledgements}

This work is supported in part by the Fund for Scientific
Research-FNRS (Belgium), by the Belgian Federal Science Policy Office
through the Interuniversity Attraction Pole P6/11, by IISN-Belgium, by
``Communaut\'e fran\c caise de Belgique - Actions de Recherche
Concert\'ees'' and by Fondecyt Projects No.~1085322 and
No.~1090753. 

\appendix

\section{Evaluation of the surface charge one-forms}
\label{sec:comp-surf-charge}

When evaluated at a spherical cross-section of $\scri$, the surface
charge one-forms \eqref{eq:29} become 
\begin{multline}
\ndelta  \cQ_\xi[h, g]= \frac{1}{16 \pi G}\lim_{r \to \infty} \int d^2
  \Omega^\varphi \ r^2 e^{2 \beta}\Big[\xi^r( D^u h- D_\sigma
  h^{u\sigma} + D^r h^u_r- D^u h^r_r)\\-\xi^u( D^r h-
  D_\sigma h^{r\sigma}- D^r h^u_u+ D^u h^r_u)+\xi^A( D^r
  h^{u}_A- D^u h^{r}_A) +\frac{1}{2}h(
  D^r\xi^u- D^u\xi^r)\\+\half h^{r\sigma}( D^u\xi_\sigma-
  D_\sigma\xi^u)-\half  h^{u\sigma}( D^r\xi_\sigma-
  D_\sigma\xi^r)\Big].\label{eq:intch} 
\end{multline}
Using the Christoffel symbols for a metric of the form \eqref{eq:2},
explicitly given in section {\bf 4.3} of \cite{Barnich:2010eb} and the solution to the
equations of motion up to the appropriate order as summarized in
section \bref{sec:solution-space}, we have
\begin{equation}
\begin{split}
 D^u h- D_\sigma
  h^{u\sigma} + D^r h^u_r- D^u h^r_r
 & =  g^{ur}g^{AB}( D_r h_{AB}-D_A h_{rB})\\
&\hspace*{-2cm}= - e^{-2 \beta} \left( g^{AB} \partial_r h_{AB} -
  k^{AB}h_{AB}+ e^{-2\beta} g^{AB} k_{AB} h_{ru}\right)  \\
&=\frac{1}{4r^3} C^{AB} \delta C_{AB} + o(r^{-3-\epsilon})\,,
\end{split}
\end{equation} 
\begin{equation}
\begin{split}
  -\Big( D^r h- D_\sigma h^{r\sigma} & - D^r h^u_u+ D^u h^r_u\Big) =
  D^A h^r_A- D^r h^A_A\\
  &= g^{ur}g^{AB}( D_A h_{uB}- D_u h_{AB}) + O(r^{-3})\\
  &=g^{ur}\left( g^{AB}\, {}^{(2)}D_B h_{uA} - h_{ur} g^{AB}(l_{AB} +
    k_{AB}\frac{V}{r})\right.  \\ & \qquad \qquad - k h_{uu} - g^{AB} \partial_u h_{AB} + g^{AB}
  h_{CA} l^C_B\Big)+ O(r^{-3}) \\ 
  &=\frac{1}{r^2} \left(4 \delta M - \frac{1}{2} \bar D_A\bar D_B \delta C^{AB} +
    \frac{1}{2} \delta \partial_u (C^{AB}C_{AB}) \right.  \\ & \qquad\qquad
  \left. - \frac{1}{2} \partial_u
    C_{AB} \delta C^{AB} - C^{AB} \partial_u \delta C_{AB} \right) + 
o(r^{-2-\epsilon})\,,
\end{split}
\end{equation}
\begin{equation}
\begin{split}
  D^r h^u_A- D^u h^r_A & = (g^{ur})^2\left( \Gamma^C_{rA}h_{uC} - 
    \partial_r h_{Au}\right) + g^{ur} g^{rB}
  \left( \Gamma^C_{rB}h_{AC} - \partial_r h_{AB}\right) + O(r^{-3}) \\
  & = \frac{1}{2 r} \bar D_B \delta C^B_A + \frac{2}{3r^2} (2 \text{
    ln }r - \frac{1}{3}) \bar D_B \delta D^B_A + o(r^{-2-\epsilon}) \\
 &  \quad  + \frac{1}{r^2} \left( \frac{4}{3}\delta N_A +
    \frac{1}{3} \delta (C_{AB} \bar D_C C^{BC}) - \frac{1}{4} C_{AB}
    \bar D_C \delta C^{BC}\right)\,,
\end{split}
\end{equation}
\begin{multline}
\half h^{r\sigma}( D^u\xi_\sigma-
  D_\sigma\xi^u)-\half  h^{u\sigma}( D^r\xi_\sigma-
  D_\sigma\xi^r)=\\
\frac{1}{2} \left(h^u_u + h^r_r \right)\left(D^u \xi^r - D^r \xi^u \right)+ 
\frac{1}{2} h^r_A\left(D^u \xi^A - D^A \xi^u \right)\,,
\end{multline}
\begin{equation}
\frac{1}{2}(h-h^u_u - h^r_r )(
  D^r\xi^u- D^u\xi^r)  = \frac{1}{2}g^{AB}h_{AB}(
  D^r\xi^u- D^u\xi^r) = 0\,,
\end{equation}
\begin{equation}
\begin{split}
D^u \xi^A - D^A \xi^u & = g^{ur} \partial_r \xi^A - g^{AB} \partial_B
\xi^u + (g^{ur} \Gamma^A_{rC} - g^{AB}\Gamma^u_{BC})\xi^C+O(r^{-3}) \\
&= \frac{-2}{r} Y^A + \frac{1}{r^2} C^A_CY^C+O(r^{-3})\,,
\end{split}
\end{equation}
\begin{multline}
\frac{1}{2}h^r_A  =  -\frac{1}{4} \bar D_B \delta C^B_A - \frac{1}{3r}
(\text{ln } r + \frac{1}{3}) \bar D_B \delta D^B_A  \\ + \frac{1}{r}
\left(-\frac{1}{3} \delta N_A - \frac{1}{12} \delta (C_{AB} \bar D_C C^{BC})
  + \frac{1}{4} \delta C_{AB} \bar D_C C^{BC} \right)+o(r^{-1-\epsilon})\,.
\end{multline}
Putting everything together, we get
\begin{multline}
  \label{eq:36bis}
 \ndelta   \cQ_\xi[\delta\cX,\cX]=\frac{1}{16 \pi G}\lim_{r \to \infty} \int
   d^2 \Omega^\varphi \,
\Big[ r \left( Y^A \frac{1}{2 } \bar D_B \delta C^B_A + Y^A
  \frac{1}{2 } \bar D_B \delta C^B_A  \right)  \\
 + Y^A \bar D_B \delta D^B_A \left(\frac{4}{3} \text{ ln } r -
  \frac{2}{9} +\frac{2}{3} \text{ ln } r + \frac{2}{9}
\right) -\frac{\psi}{8} C^{AB} \delta C_{AB}\\
+ f \left(4 \delta M - \frac{1}{2} \bar D_A\bar D_B \delta C^{AB} +
\frac{1}{2} \delta \partial_u (C^{AB}C_{AB})  - \frac{1}{2} \partial_u
C_{AB} \delta C^{AB} - C^{AB} \partial_u \delta C_{AB}\right)\\ 
 +Y^A \left( \frac{4}{3}\delta N_A + \frac{1}{3} \delta (C_{AB}
  \bar D_C C^{BC}) -
  \frac{1}{4} C_{AB} \bar D_C \delta C^{BC}\right)  \\ 
 -2 Y^A \left(-\frac{1}{3} \delta N_A - \frac{1}{12} \delta
  (C_{AB} \bar D_C C^{BC})
  + \frac{1}{4} \delta C_{AB} \bar D_C C^{BC} \right)  \\ 
 -\frac{1}{2} \bar D_A f \bar D_B \delta C^{AB} -\frac{1}{4}
C_{AB} Y^A \bar D_C
\delta C^{BC}\Big]. 
\end{multline}
Using integrations by parts and the conformal Killing equation for the
$Y^A$, this can be simplified to
\begin{equation}
\begin{split}
  \label{eq:36ter}
  \ndelta  \cQ_\xi[\delta\cX,\cX]&=\frac{1}{16 \pi G}\int
   d^2 \Omega^\varphi \,
\Big[ -\frac{\psi}{8} C^{AB} \delta C_{AB} +Y^A 2\delta N_A
-\frac{1}{2} \bar D_A f \bar D_B \delta C^{AB}\\
& \qquad\qquad + f \left(4 \delta M - \frac{1}{2} \bar D_A\bar D_B \delta C^{AB} +
\frac{1}{2} \delta \partial_u (C^{AB}C_{AB}) \right.  \\ &
\qquad\qquad \left. - \frac{1}{2} \partial_u
C_{AB} \delta C^{AB} - C^{AB} \partial_u \delta C_{AB} \right)\Big]\\ &=\frac{1}{16 \pi G}\delta \int
   d^2 \Omega^\varphi \,
\Big[ -\frac{\psi}{16} C^{AB} C_{AB} + 2Y^A N_A+ 4 f M \Big]  \\
&\qquad\qquad +\frac{1}{16 \pi G}\int
   d^2 \Omega\,
\Big[ \frac{f}{2} \partial_u
C_{AB} \delta C^{AB}\Big]. 
\end{split}
\end{equation}

\section{Computation of the charge algebra}
  \label{sec:comp-charge-algebra}

We will start by computing the usual factor,
\begin{multline}
-\delta_{s_2} Q_{s_1}[\cX]=\frac{1}{16 \pi G}\int
   d^2 \Omega^\varphi\,
\Big[ Y_1^A \left(2 (-\delta_{s_2}) N_A + \frac{1}{16}\partial_A
  (-\delta_{s_2}) (C^{CB} C_{CB})\right)\\+ 4 f_1 (-\delta_{s_2}) M \Big]\,,
\end{multline}
and organize according to the different types of terms that appear:
\begin{itemize}

\item terms containing $M$
\begin{eqnarray}
-\delta_{s_2} Q_{s_1}[\cX]_M&=&\frac{1}{16 \pi G}\int
   d^2 \Omega^\varphi\,
\Big[ Y_1^A 2 (f_2 \partial_A M + 3 \partial_A f_2 M) + 4 f_1
(Y_2^A\d_AM+\frac{3}{2} \psi_2M)\Big]\nonumber\\
&=&\frac{1}{16 \pi G}\int
   d^2 \Omega^\varphi\, 4M
\Big[ - \half \bar D_A ( Y_1^A f_2) + \frac{3}{2} Y^A_1 \partial_A f_2
- \bar D_A (f_1
Y_2^A)+\frac{3}{2} f_1\psi_2)\Big]\nonumber\\
&=&\frac{1}{16 \pi G}\int
   d^2 \Omega^\varphi\, 4M
\Big[ Y^A_1 \partial_A f_2  - \half \psi_1 f_2  -Y_2^A\partial_A f_1
+\frac{1}{2} f_1\psi_2)\Big]\nonumber\\
&=&\frac{1}{16 \pi G}\int
   d^2 \Omega^\varphi\, 4M f_{[s_1,s_2]}\,,
\end{eqnarray}

\item terms containing $N_A$
\begin{eqnarray}
-\delta_{s_2} Q_{s_1}[\cX]_N&=&\frac{1}{16 \pi G}\int
   d^2 \Omega^\varphi\,
\Big[ 2Y_1^A (\cL_{Y_2} + \psi_2) N_A \Big]\nonumber \\
&=&\frac{1}{16 \pi G}\int
   d^2 \Omega^\varphi\,
\Big[ 2Y_1^A (Y_2^B \bar D_B + \psi_2) N_A + 2 Y_1^A \bar D_A Y_2^B N_B \Big]\nonumber \\
&=&\frac{1}{16 \pi G}\int
   d^2 \Omega^\varphi\, 2 N_A
\Big[ -Y_2^B \bar D_B  Y_1^A + Y_1^B \bar D_B Y_2^A\Big]\nonumber \\
&=&\frac{1}{16 \pi G}\int
   d^2 \Omega^\varphi\, 2 N_A Y^A_{[s_1,s_2]}\,,
\end{eqnarray}

\item terms containing $D_{AB}$
\begin{eqnarray}
-\delta_{s_2} Q_{s_1}[\cX]_D&=&\frac{1}{16 \pi G}\int
   d^2 \Omega^\varphi\, 2 Y_1^A 
\Big[  -\half [\bar D_B \psi_2 +\psi_2
  \bar D_B ] D^B_A\Big]\nonumber \\
&=&\frac{1}{16 \pi G}\int
   d^2 \Omega^\varphi\, 2 Y_1^A 
\Big[ -\half \bar D_B(\psi_2 D^B_A)\Big]\nonumber \\
&=&\frac{1}{16 \pi G}\int
   d^2 \Omega^\varphi\, \bar D^B Y_1^A \psi_2 D_{AB}=0\,,
\end{eqnarray}

\item terms containing the news
\begin{eqnarray}
  -\delta_{s_2} Q_{s_1}[\cX]_{news}&=&\frac{1}{16 \pi G}\int
  d^2 \Omega^\varphi\,
  \Big[ 2Y_1^A \left( -\frac{3}{16} \bar D_A f_2
    N^B_CC^C_B
    +\half \bar D_B f_2 N^B_C C^C_A\right) \nonumber \\ && \qquad +2 Y^A_1
  f_2 \left( \frac{1}{16}\d_A\big[N^B_C C^C_B\big]
    -\frac{1}{4} \bar D_AC^C_BN^B_C
    -\frac{1}{4} \bar D_B\big[C^{B}_{C}N^C_{A}-N^B_CC^C_A\big]\right)
  \nonumber 
\\ && \qquad - \psi_1\frac{1}{8}
  C^{AB} f_2 N_{AB}+ 4 f_1 \left(\frac{1}{4} \bar D_B \bar D_C f_2 N^{BC} + \half
    \bar D_B f_2 \bar D_C N^{BC} \right)\nonumber \\ && \qquad+ 4 f_1
  f_2 
\left( -\frac{1}{8} N^A_BN^B_A
    +\frac{1}{4}\bar D_A \bar D_C N^{CA} \right)\Big] \nonumber \\
  &=&\frac{1}{16 \pi G}\int
  d^2 \Omega^\varphi\,\frac{-1}{2} N^{BC} f_2 
  \Big[ f_1 N_{BC} + \cL_{Y_1} C_{BC} - \half \psi_1 C_{BC} -2 \bar D_B \bar D_C
  f_1\Big]\nonumber \\ && + \frac{1}{16 \pi G}\int
  d^2 \Omega^\varphi\,\half N^{C}_B C_{CA} \Big[ Y_1^A \bar D^Bf_2 +
  Y^B_1 \bar D^A f_2 - \bar \gamma^{AB} Y_1^D \bar D_Df_2\Big]\,.
\end{eqnarray}
The second line is zero. This is coming from the following identity
for the symmetrized product of two traceless matrices in 2
dimensions,
\begin{equation}
\half (C^{A}_B K^{B}_C+K^A_B C^B_C) = \half \delta^{A}_{C} C^B_D K^D_B\,,
\end{equation}
and the conformal Killing equation for the $Y^A$. The first line can
be recognized as, 
\begin{eqnarray}
-\delta_{s_2} Q_{s_1}[\cX]_{news}&=&\frac{1}{16 \pi G}\int
   d^2 \Omega^\varphi\,\frac{-1}{2} N^{BC} f_2 
\Big[-\delta_{s_1} C_{BC} \Big]\nonumber \\
&=&- \Theta_{s_2} [-\delta_{s_1} \cX,\cX] \,,
\end{eqnarray}

\item the rest
\begin{eqnarray}
  -\delta_{s_2} Q_{s_1}[\cX]_R&=&\frac{1}{16 \pi G}\int
  d^2 \Omega^\varphi\,
  \Big[ 2Y_1^A  \Big( -\frac{1}{32}\bar D_A\psi_2
  C^B_CC^C_B +f_2 \frac{1}{4}C_A^B\d_B
  \bar  R\nonumber\\ &&\qquad -\frac{1}{4}f_2
  \bar D_B \left( \bar D^B \bar D_CC^C_A -\bar D_A 
    \bar D_CC^{BC}\right)\nonumber\\ &&\qquad + \frac{1}{4} ( \bar
  D_Bf_2 \bar R+\bar 
  D_B \bar \Delta
  f_2)C^B_A -\frac{3}{4} \bar D_B f_2(\bar D^B\bar D_C C^C_A-
  \bar D_A \bar D_C C^{BC})
  \nonumber\\ &&\qquad +\half (\bar D_A\bar D_B f_2
  -\frac{1}{2} \bar \Delta f_2\bar \gamma_{AB} ) \bar D_C C^{CB}+\frac{3}{8} 
  \bar D_A(\bar D_C\bar D_B f_2 C^{CB}) \Big) \nonumber \\ && \qquad -\psi_1 \frac{1}{8}
  C^{CB} \left( [\cL_{Y_2} -\half\psi_2] C_{CB}-2 \bar D_C \bar D_B f_2+
    \bar \Delta f_2\bar 
\gamma_{CB}\right)\nonumber \\ && \qquad+ 4 f_1 \left(f_2 \frac{1}{8}
    \bar \Delta  \bar R +\frac{1}{4}\d_A f_2 \d^A \bar R +\frac{1}{8}
    \bar D_C \bar D_B \psi_2 C^{CB} \right) \Big]\nonumber \\ 
  &=&\frac{1}{16 \pi G}\int
  d^2 \Omega^\varphi\,
  \Big[ -Y_1^A \frac{1}{16}\bar D_A\psi_2
  C^B_CC^C_B -\psi_1 \frac{1}{8}
  C^{CB} \left( [\cL_{Y_2} -\half\psi_2] C_{CB}\right) \nonumber\\
  &&\qquad +C^{BC} \Big( \half f_1 \bar D_B \bar D_C \psi_2 +\psi_1
  \frac{1}{4} \bar D_C\bar D_B f_2 + \half f_2 Y_{1B} \d_C \bar R \nonumber\\
  &&\qquad  -\frac{3}{4} \psi_1 \bar D_B \bar D_C f_2 - \bar D_C
  (Y_1^A \bar D_A\bar 
  D_B f_2)
  +\frac{1}{2} \bar D_C ( Y_{1B}\bar \Delta f_2 ) \nonumber\\ &&\qquad
  +\frac{1}{2} Y_{1C}( \bar D_Bf_2\bar R+
  \bar D_B \bar \Delta
  f_2) + \half \bar D_C \bar \Delta (Y_{1B} f_2) - \half \bar D_C \bar D_A \bar D_B (Y_1^A f_2)
  \nonumber\\ &&\qquad -\frac{3}{2} \bar D_C \bar D_A (Y_{1B} \bar D^Af_2) +
  \frac{3}{2} \bar D_C \bar D_A (Y_1^A D_B f_2)\Big) \nonumber\\ &&\qquad+\half ( f_1
  \d_A f_2 - f_2 \d_A f_1)\d^A \bar R\Big]\,.
\end{eqnarray}
Using the commutation rule for covariant derivatives, this gives
\begin{eqnarray}
-\delta_{s_2} Q_{s_1}[\cX]_C&=&\frac{1}{16 \pi G}\int
   d^2 \Omega^\varphi\,
\Big[ -\frac{1}{16}(Y_1^A  \bar D_A\psi_2-Y_2^A  \bar D_A\psi_1)
  C^B_CC^C_B \nonumber\\ && \qquad+\half ( f_1
  \d_A f_2 - f_2 \d_A f_1)\d^A \bar R
   +C^{BC} \Big(\half (f_1 \bar D_B \bar D_C \psi_2 - f_2 \bar D_B \bar D_C
  \psi_1) \nonumber \\ && \qquad
+ \frac{1}{4}f_2 Y_{1B} \d_C \bar R + \half f_2 \bar D_C \bar \Delta Y_{1B}
   + \frac{1}{4}\bar D_Cf_2 Y_{1B} \bar R + \half
  \bar D_C f_2 \bar \Delta Y_{1B}\Big) \Big]\nonumber\\
&=&\frac{1}{16 \pi G}\int
   d^2 \Omega^\varphi\,
\Big[ -\frac{1}{16}\psi_{[s_1,s_2]}
  C^B_CC^C_B +\half ( f_1
  \d_A f_2 - f_2 \d_A f_1)\d^A \bar R\nonumber\\
  &&\qquad +C^{BC} \half (f_1 \bar D_B \bar D_C \psi_2 - f_2 \bar D_B \bar D_C
  \psi_1) \Big]\,,
\end{eqnarray}
where in the last line we have used the identity $\bar \Delta Y^A =
-\half \bar R Y^A$ satisfied by conformal Killing vectors.

\end{itemize}

Summing everything, we obtain
\begin{eqnarray}
-\delta_{s_2} Q_{s_1}[\cX]
&=&\frac{1}{16 \pi G}\int
   d^2 \Omega^\varphi\,
\Big[ -\frac{1}{16}\psi_{[s_1,s_2]}
  C^B_CC^C_B +\half ( f_1
  \d_A f_2 - f_2 \d_A f_1)\d^A \bar R\nonumber\\
  &&\qquad +C^{BC} \half (f_1 \bar D_B \bar D_C \psi_2 - f_2 \bar D_B
  \bar D_C
  \psi_1) +4M f_{[s_1,s_2]}+2 N_A Y^A_{[s_1,s_2]}\Big]\nonumber \\
&& \qquad- \Theta_2 [-\delta_1 \cX,\cX] \nonumber \\
&=& Q_{[s_1,s_2]}- \Theta_2 [-\delta_1 \cX,\cX] +K_{s_1,s_2}[\cX]\,,
\end{eqnarray}
with $K_{s_1,s_2}[\cX]$ defined in \eqref{centralcha}.

\section{Checking the cocyle condition}
\label{sec:verif-cocyle-cond}

Let us treat the two parts of $K_{s_1,s_2}[\cX]$ separately:
\begin{itemize}

\item for the second part $\hat K_{s_1,s_2}= \frac{1}{16 \pi G}\int
   d^2 \Omega^\varphi\,C^{BC} \half (f_1 \bar D_B \bar D_C \psi_2 -
   f_2 \bar D_B \bar D_C
  \psi_1) $, we have
\begin{eqnarray}
  A&=&\int
  d^2 \Omega^\varphi\,\Big[(-\delta_{s_3} C^{BC})  (f_1 \bar D_B \bar
  D_C \psi_2 - f_2 \bar D_B \bar D_C
  \psi_1) +{\rm cyclic}\ (1,2,3) \Big]\nonumber \\
  &=&\int
  d^2 \Omega^\varphi\,\Big[([f_3\d_u+\cL_{Y_3} -\half \psi_3] C_{AB}-2
  \bar D_A\bar D_B f_3+\Delta f_3\bar \gamma_{AB})\nonumber \\ && \qquad
  \qquad \qquad  (f_1 \bar D^B \bar D^A \psi_2 - f_2 \bar D^B \bar D^A
  \psi_1) +{\rm cyclic}\ (1,2,3)\Big]\nonumber \\
  &=&\int
  d^2 \Omega^\varphi\,\Big[-C_{BC} \bar D_A \left((Y_1^A f_2 - Y_2^A
    f_1) \bar D^B \bar D^C \psi_3 \right)\nonumber \\ && 
  +2 C_{BC} \left(\bar D_A Y^B_1 f_2 -\bar D_A Y^B_2 f_1
  \right) \bar D^A
  \bar D^C \psi_3  -\half C_{BC} \left(\psi_1 f_2 -\psi_2 f_1 \right) \bar D^B
  \bar D^C \psi_3\nonumber \\ && 
  +2 \left(\bar D_C f_1 f_2 - \bar D_C f_2 f_1 \right) \left(
    \bar \Delta \bar D^C
    \psi_3 - \half \bar D^C \bar \Delta \psi_3\right)+{\rm cyclic}\ (1,2,3) \Big],
\end{eqnarray}
The second term is given by
\begin{eqnarray}
  B&=&\int
  d^2 \Omega^\varphi\,\Big[C^{BC}  (f_{[s_1,s_2]} \bar D_B \bar D_C
  \psi_3 - f_3 \bar D_B \bar D_C
  \psi_{[s_1,s_2]})+{\rm cyclic}\ (1,2,3) \Big]\nonumber \\
  &=&\int
  d^2 \Omega^\varphi\,C^{BC} \Big[\bar D_A \left( (Y^A_1 f_2 - Y^A_2
    f_1)\bar  D_B \bar D_C
    \psi_3\right)  - \frac{3}{2} \left( \psi_1
    f_2 - \psi_2 f_1\right) \bar D_B \bar D_C \psi_3 \nonumber \\ &&  -
  \left(Y^A_1 f_2 - Y^A_2 f_1 \right) \bar D_A \bar D_B \bar D_C
  \psi_3
- f_3 \bar D_B \bar D_C \left(Y^A_1 \bar D_A \psi_2
    - Y^A_2 \bar D_A \psi_1 \right)+{\rm cyclic}\ (1,2,3) \Big]\nonumber \\
  &=&\int
  d^2 \Omega^\varphi\,C^{BC} \Big[\bar D_A \left( (Y^A_1 f_2 - Y^A_2
    f_1) \bar D_B \bar D_C
    \psi_3\right) - \frac{3}{2} \left( \psi_1
    f_2 - \psi_2 f_1\right) \bar D_B \bar D_C \psi_3 \nonumber \\ && \qquad
  -2 \left( f_1 \bar D_B Y^A_2 - f_2 \bar D_B Y^A_1\right) \bar D_C
  \bar D_A \psi_3+{\rm cyclic}\ (1,2,3)\Big]\,.
\end{eqnarray}
Summing the two, we get
\begin{eqnarray}
A+B & = & \int
   d^2 \Omega^\varphi\,\Big\{C^{BC} \Big[-2 \left( f_1 (\bar D_B Y^A_2 +
     \bar D^A Y_{2B}) - f_2 (\bar D_B Y^A_1 + \bar D^A
       Y_{1B})\right) \bar D_C \bar D_A \psi_3 \nonumber\\ && - 2 \left( \psi_1
       f_2 - \psi_2 f_1\right) \bar D_B \bar D_C \psi_3 \Big]+2 \left(\bar
       D_C f_1 f_2 - \bar D_C f_2 f_1 \right) \left( \bar \Delta \bar D^C
   \psi_3 - \half \bar D^C \bar \Delta \psi_3\right)\nonumber \\
     &&
\qquad \qquad +{\rm cyclic}\ (1,2,3) \Big\}\nonumber \\ && \hspace*{-2cm}
=\int
   d^2 \Omega^\varphi\,\Big[ 2 \left(\bar D_C f_1 f_2 - \bar D_C f_2
     f_1 \right) 
\left( \bar \Delta \bar D^C
   \psi_3 - \half \bar D^C \bar \Delta \psi_3\right)+{\rm cyclic}\ (1,2,3) \Big]\,.
\end{eqnarray}
We can then use the following identities $ \bar \Delta \psi = -\bar
D_A(\bar R Y^A)$ and $\Delta \bar D^C \psi = \bar D^C \Delta\psi +
\half \bar R \bar D^C \psi$ that can be deduced from the identity
(4.59) for covariant derivatives of conformal Killing vectors in
\cite{Barnich:2010eb}\footnote{Note that the first identity corrects
  the corresponding identity of \cite{Barnich:2010eb} in the case of
  non constant curvature. Note also that the second relation after
  (4.57) in \cite{Barnich:2010eb} should be replaced by $ \bar D_A f
  \bar D_C C^C_B+\bar D_B f \bar D_C C^C_A+ \bar D_C f \bar D_A C^C_B+
  \bar D_C f \bar D_B C^C_A- 2\bar D^C f\bar D_C
  C_{AB}-2\bar\gamma_{AB} \bar D_C f\bar D_D C^{CD} =0$.}  to simplify
the above to
\begin{eqnarray}
A&+&B  = \int
   d^2 \Omega^\varphi\,\Big[ 2 \left(\bar D_C f_1 f_2 - \bar D_C f_2 f_1 \right) \left(
     -\half \bar D^C (Y_3^A \bar D_A R) - \psi_3\half \bar D^C \bar R
   \right)+{\rm cyclic}\ (1,2,3) 
   \Big]\nonumber \\ \hspace*{-4cm}
&=&\int
   d^2 \Omega^\varphi\,\Big[ 2 \left(\bar D_C f_1 f_2 - \bar D_C f_2 f_1 \right) \left(
     -\half \cL_{Y_3} \bar D^C\bar R - \psi_3\bar D^C \bar R \right)+{\rm cyclic}\ (1,2,3)
   \Big]\,.
\end{eqnarray}

\item for the first part $\tilde K_{s_1,s_2}= \frac{1}{16 \pi G}\int
   d^2 \Omega^\varphi\, \half ( f_1
  \d_A f_2 - f_2 \d_A f_1)\d^A \bar R $, condition (\ref{cocycle})
  leads to 
\begin{eqnarray}
C & = & \int d^2 \Omega^\varphi\,\Big[ f_{[s_1,s_2]} \d_A f_3 - f_3
\d_A f_{[s_1,s_2]} \d^A \bar R
+{\rm cyclic}\ (1,2,3)\Big]\nonumber \\ 
&=& \int d^2 \Omega^\varphi\,\Big[ \cL_{Y_1} (f_2 \d_A f_3- f_3 \d_A f_2 ) -
\psi_1 (f_2 \d_A f_3- f_3 \d_A f_2 )\d^A \bar R+
{\rm cyclic}\ (1,2,3)\Big]\nonumber \\ 
&=& \int d^2 \Omega^\varphi\,\Big[ (f_2 \d_A f_3- f_3 \d_A f_2 ) (-\cL_{Y_1} -
2 \psi_1)\d^A \bar R +{\rm cyclic}\ (1,2,3)\Big]\,.
\end{eqnarray}

\end{itemize}

The different contributions then sum up to zero, $A+B+C=0$. 

\section{Kerr solution in BMS gauge}
\label{sec:kerr-solution-bms}

We start from equation (48) of \cite{0264-9381-20-19-302} giving
the Kerr metric in generalized Bond-Metzner-Sachs coordinates, that is
to say in a coordinate system $u,\tilde r,\theta,\phi$ such that
$g_{\tilde r\tilde r}=g_{\tilde r A}=0$. When changing the signature
to $(-,+,+,+)$ and expanding in $\tilde r$, one finds

\begin{eqnarray}
  \label{eq:15}
  g_{uu}&=&-1 + 2 M\,{\tilde r}^{-1}+ O(\tilde r^{-2})\,,\\
g_{u\tilde r}&=&-1+a^2(\frac{1}{2}-\cos^2{\theta})\,\tilde r^{-2}+O(\tilde
r^{-3})\,,\\
g_{u\theta}&=&-a\cos{\theta}+2a\cos\theta(M-a\sin\theta)\,\tilde r^{-1}+
O(\tilde r^{-2})\,,\\
g_{u\phi}&=&-2aM\sin^2{\theta}\, \tilde r^{-1}+
O(\tilde r^{-2})\,,\\
g_{\theta \theta} & = & \tilde r^2 + 2 a \sin \theta \, \tilde r + a^2
(3 \sin^2 \theta - 1) + O(\tilde r^{-1})\,,\\
g_{\phi \phi} & = & \tilde r^2 \sin^2 \theta - 2 a \sin \theta \cos^2
\theta \, \tilde r+ a^2 (1- 3 \sin^2 \theta \cos^2 \theta)+ O(\tilde r^{-1})\\
g_{\theta \phi}& = & O(\tilde r^{-1})\,.
\end{eqnarray}
The Bondi-Metzner-Sachs gauge is reached by defining $r$ through ${\rm
  det}\, g_{AB}= r^4 \sin^2\theta$, which implies that 
\begin{equation}
  \label{eq:23}
\tilde  r =  r +\frac{a}{2} \frac{\cos (2 \theta)}{\sin \theta} +
 \frac{a^2}{8 } (4 \cos (2 \theta) + \frac{1}{\sin^2 \theta})\, 
 r^{-1} + O( r^{-2})\,.
\end{equation}
In the coordinates $u,r,\theta,\phi$, the metric components
$g_{uu},g_{ur},g_{u\phi}$ are simply obtained from the
above expressions by replacing $\tilde r$ by $r$, while 
\begin{eqnarray}
g_{\theta \theta} & = & r^2 +  \frac{a}{\sin \theta} \, r + \frac{a^2}{2 \sin^2 \theta} + O(r^{-1})\,,\\
g_{\phi \phi} & = & r^2 \sin^2 \theta - a \sin \theta \, r+ \frac{a^2}{2}+ O(r^{-1})\,,\\
g_{\theta \phi}& = & O(r^{-1})\,,\\
g_{u\theta} &=&\frac{a}{2}  \frac{\cos \theta}{\sin^2
  \theta}+\frac{a\cos\theta }{4}
(8 M + \frac{a}{\sin^3 \theta}) \,
r^{-1} +O(r^{-2})\,.
\end{eqnarray}
When comparing with section \bref{sec:solution-space}, one can read
off $\cX^{Kerr}$ as described at the beginning of section
\ref{sec:centr-extens-kerr}.  

\section{Integration on the sphere}
\label{sec:formulas-sphere}

Consider stereographic coordinates
$\zeta=e^{i\phi}\cot{\frac{\theta}{2}}$ and let $\mu=\cos\theta$,
$P=\half(1+\zeta\bar\zeta)$. We have
\begin{equation}
\begin{gathered}
  \label{eq:1}
  \int^{2\pi}_0 d\phi\int^\pi_0\sin\theta d\theta\,
  \zeta^m\zeta^n=4\pi\delta_{m+n}^0,\quad 
\int^{2\pi}_0 d\phi\int^\pi_0\sin\theta d\theta\,
  \bar\zeta^m\bar\zeta^n=4\pi\delta_{m+n}^0,\\
\int^{2\pi}_0 d\phi\int^\pi_0\sin\theta d\theta\,
  \zeta^m\bar\zeta^n=2\pi\delta_{m}^n\int^1_{-1}d\mu
\big(\frac{1+\mu}{1-\mu}\big)^m,
\end{gathered}
\end{equation}

$d^2\Omega=\sin\theta d\theta\wedge d\phi=\frac{d\zeta\wedge
  d\bar\zeta}{2iP^2}$, 

\begin{equation}
\begin{gathered}
  \label{eq:51}
  \cos{\theta}=-\frac{1-\zeta\bar\zeta}{1+\zeta\bar\zeta}=\mu,\quad 
\sin\theta=P^{-1}\sqrt{\zeta\bar\zeta},\\ 
2\sin^2{\frac{\theta}{2}}=P^{-1}=1-\mu,\quad
\zeta\bar\zeta=\frac{1+\mu}{1-\mu},
\end{gathered}
\end{equation}

\begin{eqnarray}
  \label{eq:28}
 \frac{\d(\zeta,\bar\zeta)}{\d(\theta,\phi)}=\left( \begin{array}{cc}
     -P\sqrt{\frac{\zeta}{\bar\zeta}} & i\zeta\\
-P\sqrt{\frac{\bar \zeta}{\zeta}} & -i\bar\zeta
\end{array}\right)\,,\qquad  \frac{\d(\theta,\phi)}{\d(\zeta,\bar\zeta)}=\left( \begin{array}{cc}
     -\frac{1}{2P}\sqrt{\frac{\bar\zeta}{\zeta}} &  -\frac{1}{2P}\sqrt{\frac{\zeta}{\bar\zeta}}\\
-\frac{i}{2\zeta} & \frac{i}{2\bar\zeta}
\end{array}\right) \,.
\end{eqnarray}

\section{Computations for the Kerr black hole}
\label{sec:kerr-bms-coordinates}

\begin{equation}
\begin{gathered}
\bar D_B C^{B\theta}=\frac{a\cos{\theta}}{\sin^2{\theta}},\quad \bar
D_B C^{B\phi}=0,\quad
C^{\zeta\zeta}=
\frac{a}{8}\frac{(1+\zeta\bar\zeta)^32\zeta^2}{\sqrt{\zeta^3\bar\zeta^3}},\\
D_\zeta
C^{\zeta\zeta}=\frac{a}{8}\frac{(1+\zeta\bar\zeta)^2(\zeta-
\zeta^2\bar\zeta)}{\sqrt{\zeta^3\bar\zeta^3}},\quad 
D_\zeta D_\zeta
C^{\zeta\zeta}=-\frac{a}{16}\frac{(1+\zeta\bar\zeta)^3}{\sqrt{\zeta^3\bar\zeta^3}},
\\
C^{AB}C_{AB}=\frac{2a^2}{\sin^2{\theta}}=
\frac{a^2}{2}\frac{(1+\zeta\bar\zeta)^2}{\zeta\bar\zeta}\label{eq:52}\,.
\end{gathered}
\end{equation}
For $Y_m=-\zeta^{m+1}$,
\begin{equation}
\begin{gathered}
\psi_m=\frac{-(m+1)\zeta^m+(1-m)\zeta^{m+1}\bar\zeta}{1+\zeta\bar\zeta},\\
\d\psi_m=\frac{-m(m+1)\zeta^{m-1}+2(1-m^2)\zeta^m\bar\zeta+m(1-m)\zeta^{m+1}
\bar\zeta^2}{(1+\zeta\bar\zeta)^2},\\ 
\bar\d\psi_m=\frac{2\zeta^{m+1}}{(1+\zeta\bar\zeta)^2}\,.\label{eq:53}
\end{gathered}
\end{equation}
\begin{equation}
\begin{gathered}
N_\zeta=\frac{4}{(1+\zeta\bar\zeta)^2}
\Big[\frac{3aM}{2}(\half\sqrt{\frac{\bar\zeta}{\zeta}}(1-\zeta\bar\zeta)+i\bar\zeta) 
+\frac{a^2}{256}(1-\zeta\bar\zeta)(1+\zeta\bar\zeta)^3\frac{1}{\zeta^2\bar\zeta}
\Big],\\
N_\zeta+\frac{1}{32}\d(C^{AB}C_{AB})=\frac{4}{(1+\zeta\bar\zeta)^2}
\Big[\frac{3aM}{2}(\half\sqrt{\frac{\bar\zeta}{\zeta}}(1-\zeta\bar\zeta)+i\bar\zeta)
\Big].
\label{eq:54}
\end{gathered}
\end{equation}
\begin{equation}
C^{AB} D_A D_B \psi = C^{\zeta\zeta}(\d \partial \psi -\Gamma \d
\psi)+C^{\bar\zeta \bar\zeta}(\bar \d \bar\partial \psi -\bar\Gamma \bar
\d \psi),
\end{equation}
\begin{eqnarray}
\d \partial \psi_m -\Gamma \d\psi_m& =& \d \left(\frac{-m(m+1)\zeta^{m-1}+2(1-m^2)\zeta^m\bar\zeta+m(1-m)\zeta^{m+1}
\bar\zeta^2}{(1+\zeta\bar\zeta)^2}\right)\nonumber \\ && \quad + \frac{2 \bar \zeta}{1+\zeta \bar
\zeta} \frac{-m(m+1)\zeta^{m-1}+2(1-m^2)\zeta^m\bar\zeta+m(1-lm\zeta^{m+1}
\bar\zeta^2}{(1+\zeta\bar\zeta)^2}\nonumber \\
& =& \frac{\d \left[-m(m+1)\zeta^{m-1}+2(1-m^2)\zeta^m\bar\zeta+m(1-m)\zeta^{m+1}
\bar\zeta^2 \right]}{(1+\zeta\bar\zeta)^2}\nonumber \\
&=& m (1-m^2) \frac{\zeta^{m-2} + 2 \zeta^{m-1} \bar \zeta + \zeta^m
  \bar \zeta^2}{(1+\zeta\bar\zeta)^2}\nonumber \\
&=& m (1-m^2) \zeta^{m-2},
\end{eqnarray}
\begin{equation}
\bar \d \bar\partial \psi_m -\bar\Gamma \bar
\d \psi_m =\bar \d \left(\frac{2\zeta^{m+1}}{(1+\zeta\bar\zeta)^2}\right) + \frac{2 \zeta}{1+\zeta \bar
\zeta} \frac{2\zeta^{m+1}}{(1+\zeta\bar\zeta)^2} = 0.
\end{equation}

If $T=0$ then $4fM=D_A(2uM Y^A)$ and the associated term in the charge
vanishes. More directly, for $Y_m$, we get $\frac{Mu}{2\pi G}\int
d^2\Omega\, \psi_m=\frac{Mu}{G}\int^1_{-1}d\mu\, \mu=0$.  It follows
that 
\begin{equation}
  \label{eq:25}
  Q_{0,Y}[\cX^{Kerr}]=\frac{1}{8\pi G}\int
  d^2\Omega\, Y^A\big(N_A+\frac{1}{32}\d_A(C^{BC}C_{BC})\big).
\end{equation}
\begin{equation}
\begin{split}
  \label{eq:14}
  Q_{0,l_m}[\cX^{Kerr}]
&=-\frac{1}{8\pi G}\int
  d^2\Omega\frac{4}{(1+\zeta\bar\zeta)^2}
  \zeta^{m+1}\,\frac{3aM}{2}\big(\half\sqrt{\frac{\bar\zeta}{\zeta}}
(1-\zeta\bar\zeta)+i\bar\zeta\big)
\\
 &=-\frac{3aM}{8G}\delta^m_0\int^1_{-1} d\mu
  \frac{4}{(1+\zeta\bar\zeta)^2}  \big(\half
  \sqrt{\zeta\bar\zeta}(1-\zeta\bar\zeta)+i \zeta\bar\zeta\big) 
   \\
&=-\frac{3aM}{8G}\delta^m_0\int^1_{-1}d\mu
  \big[-\mu(1+\mu)^{1/2}{(1-\mu)^{1/2}}
  +i{(1+\mu)}{(1-\mu)}\big]\\
 &=-\delta^m_0\frac{iaM}{2G}.
\end{split}
\end{equation}

\begin{equation}
\begin{split}
 & K_{(0,l_m),(0,l_n)}[\cX^{Kerr}] = \frac{1}{32 \pi G} \int d^2\Omega
  \left\{ \frac{u}{2} \psi_m C^{\zeta\zeta} n (1-n^2)
    \zeta^{n-2} - (m\leftrightarrow n)\right \}\\
&  =\frac{u}{64 \pi G} \int d^2\Omega
  \frac{a}{4}\frac{(1+\zeta\bar\zeta)^3\zeta^2}{\sqrt{\zeta^3\bar\zeta^3}}\left\{
    \frac{-(m+1)\zeta^m+(1-m)\zeta^{m+1}\bar\zeta}{1+\zeta\bar\zeta} n
    (1-n^2)
    \zeta^{n-2} - (m\leftrightarrow n)\right \} \\
&  =\frac{u a}{256 \pi G} \int d^2\Omega
  \frac{(1+\zeta\bar\zeta)^2}{\sqrt{\zeta^3\bar\zeta^3}}\Big\{ -(m+1)
  n (1-n^2) \zeta^{m+n} +(1-m) n (1-n^2)\zeta^{m+n+1}\bar\zeta-
  (m\leftrightarrow n)
  \Big \} \\
&  =\frac{u a \, m (1-m^2)\delta^0_{m+n}}{128 \pi G} \int d^2\Omega
  \frac{(1+\zeta\bar\zeta)^2}{\sqrt{\zeta^3\bar\zeta^3}}(
  1-\zeta\bar\zeta)  \\
&  =\frac{u a \, m (m^2-1)\delta_{m+n,0}}{8 G} \int_{-1}^1
  d \mu \frac{\mu}{\sqrt{(1+\mu)^3(1-\mu)^3}} \\
 & =0\,.
\end{split}
\end{equation}
\begin{equation}
\begin{split}
&K_{(0,l_m),(0,\bar l_n)}[\cX^{Kerr}]= \frac{1}{32 \pi G} \int
d^2\Omega \left\{ \frac{u}{2} \psi_m C^{\bar\zeta\bar\zeta}  n (1-n^2)
  \bar\zeta^{n-2} - \frac{u}{2} \psi_{\bar n} C^{\zeta\zeta}  m (1-m^2)
  \zeta^{m-2}\right \} \\
&=\frac{u}{64 \pi G} \int
d^2\Omega
\frac{a}{4}\frac{(1+\zeta\bar\zeta)^2}{\sqrt{\zeta^3\bar\zeta^3}}\left\{
  -(m+1) n (1-n^2)
  \bar \zeta^n\zeta^m+(1-m)\zeta^{m+1} n (1-n^2)
  \bar \zeta^{n+1}\right.  \\ & \qquad \qquad \left. +(n+1)
  m (1-m^2)
 \zeta^m\bar \zeta^n-(1-n) \bar \zeta^{n+1} m (1-m^2)
 \zeta^{m+1}\right \}\nonumber \\
&= 0\,.
\end{split}
\end{equation}
\begin{equation}
\begin{split}
K_{(T_{m,n},0),(0, l_l)}[\cX^{Kerr}] &= \frac{1}{32 \pi G} \int
d^2\Omega\,  T_{m,n} C^{\zeta\zeta}  l (1-l^2)
  \zeta^{l-2} \\
&= \frac{a\,  l (1-l^2)}{64 \pi G} \int
d^2\Omega
\frac{(1+\zeta\bar\zeta)^2}{\sqrt{\zeta^3\bar\zeta^3}}
\zeta^{m+l} \bar \zeta^{n} .
\end{split}
\end{equation}

\newpage


\begin{thebibliography}{10}

\bibitem{Bondi:1962px}
H.~Bondi, M.~G. van~der Burg, and A.~W. Metzner, ``Gravitational waves in
  general relativity. 7. {W}aves from axisymmetric isolated systems,'' {\em
  Proc.\ Roy.\ Soc.\ Lond. A} {\bf 269} (1962)
21.

\bibitem{Sachs:1962wk}
R.~K. Sachs, ``Gravitational waves in general relativity. 8. {W}aves in
  asymptotically flat space-times,'' {\em Proc.\ Roy.\ Soc.\ Lond.\ A} {\bf
  270} (1962)
103.

\bibitem{Barnich:2009se}
G.~Barnich and C.~Troessaert, ``{Symmetries of asymptotically flat 4
  dimensional spacetimes at null infinity revisited},'' {\em Phys. Rev. Lett.}
  {\bf 105} (2010) 111103,
\href{http://www.arXiv.org/abs/0909.2617}{{\tt 0909.2617}}.

\bibitem{Barnich:2010eb}
G.~Barnich and C.~Troessaert, ``{Aspects of the BMS/CFT correspondence},'' {\em
  JHEP} {\bf 05} (2010) 062,
\href{http://www.arXiv.org/abs/1001.1541}{{\tt 1001.1541}}.

\bibitem{Wald:1999wa}
R.~M. Wald and A.~Zoupas, ``A general definition of conserved quantities in
  general relativity and other theories of gravity,'' {\em Phys. Rev.} {\bf
  D61} (2000) 084027,
\href{http://www.arXiv.org/abs/gr-qc/9911095}{{\tt gr-qc/9911095}}.

\bibitem{Brown:1986nw}
J.~D. Brown and M.~Henneaux, ``Central charges in the canonical realization of
  asymptotic symmetries: An example from three-dimensional gravity,'' {\em
  Commun. Math. Phys.} {\bf 104} (1986) 207.

\bibitem{Strominger:1998eq}
A.~Strominger, ``Black hole entropy from near-horizon microstates,'' {\em JHEP}
  {\bf 02} (1998) 009,
\href{http://www.arXiv.org/abs/arXiv:hep-th/9712251}{{\tt
  arXiv:hep-th/9712251}}.

\bibitem{Guica:2008mu}
M.~Guica, T.~Hartman, W.~Song, and A.~Strominger, ``{The Kerr/CFT
  Correspondence},'' {\em Phys. Rev.} {\bf D80} (2009) 124008,
\href{http://www.arXiv.org/abs/0809.4266}{{\tt 0809.4266}}.

\bibitem{Bredberg:2011hp}
I.~Bredberg, C.~Keeler, V.~Lysov, and A.~Strominger, ``{Cargese Lectures on the
  Kerr/CFT Correspondence},'' \href{http://www.arXiv.org/abs/1103.2355}{{\tt
  1103.2355}}. * Temporary entry *.

\bibitem{Barnich:2001jy}
G.~Barnich and F.~Brandt, ``Covariant theory of asymptotic symmetries,
  conservation laws and central charges,'' {\em Nucl. Phys.} {\bf B633} (2002)
  3--82,
\href{http://arXiv.org/abs/hep-th/0111246}{{\tt hep-th/0111246}}.

\bibitem{Barnich:2003xg}
G.~Barnich, ``Boundary charges in gauge theories: {U}sing {S}tokes theorem in
  the bulk,'' {\em Class. Quant. Grav.} {\bf 20} (2003) 3685--3698,
\href{http://www.arXiv.org/abs/hep-th/0301039}{{\tt hep-th/0301039}}.

\bibitem{Barnich:2007bf}
G.~Barnich and G.~Comp\`{e}re, ``{Surface charge algebra in gauge theories and
  thermodynamic integrability},'' {\em J. Math. Phys.} {\bf 49} (2008) 042901,
\href{http://www.arXiv.org/abs/0708.2378}{{\tt 0708.2378}}.

\bibitem{Barnich:2004ts}
G.~Barnich, S.~Leclercq, and P.~Spindel, ``Classification of surface charges
  for a spin 2 field on a curved background solution,'' {\em Lett. Math. Phys.}
  {\bf 68} (2004) 175--181,
\href{http://www.arXiv.org/abs/gr-qc/0404006}{{\tt gr-qc/0404006}}.

\bibitem{Abbott:1981ff}
L.~F. Abbott and S.~Deser, ``Stability of gravity with a cosmological
  constant,'' {\em Nucl. Phys.} {\bf B195} (1982)
76.

\bibitem{Fuks:1986}
D.~Fuks, {\em {Cohomology of infinite-dimensional Lie algebras}}.
\newblock Consultants Bureau, New York, 1986.

\bibitem{Sachs2}
R.~K. Sachs, ``Asymptotic symmetries in gravitational theories,'' {\em Phys.\
  Rev.} {\bf 128} (1962) 2851--2864.

\bibitem{goldberg:2155}
J.~N. Goldberg, A.~J. Macfarlane, E.~T. Newman, F.~Rohrlich, and E.~C.~G.
  Sudarshan, ``{Spin-s Spherical Harmonics and eth},'' {\em Journal of
  Mathematical Physics} {\bf 8} (1967), no.~11, 2155--2161.

\bibitem{held:3145}
A.~Held, E.~T. Newman, and R.~Posadas, ``{The Lorentz Group and the Sphere},''
  {\em Journal of Mathematical Physics} {\bf 11} (1970), no.~11, 3145--3154.

\bibitem{Penrose:1984}
R.~Penrose and W.~Rindler, {\em {Spinors and Space-Time, Volume 1: Two-spinor
  Calculus and Relativistic Fields}}.
\newblock Cambridge University Press, 1984.

\bibitem{Regge:1974zd}
T.~Regge and C.~Teitelboim, ``Role of surface integrals in the {H}amiltonian
  formulation of general relativity,'' {\em Ann. Phys.} {\bf 88} (1974)
286.

\bibitem{Brown:1986ed}
J.~D. Brown and M.~Henneaux, ``On the {P}oisson brackets of differentiable
  generators in classical field theory,'' {\em J. Math. Phys.} {\bf 27} (1986)
  489.

\bibitem{0264-9381-20-19-302}
S.~J. Fletcher and A.~W.~C. Lun, ``{The Kerr spacetime in generalized
  Bondi-Sachs coordinates},'' {\em Classical and Quantum Gravity} {\bf 20}
  (2003), no.~19, 4153--4167.

\bibitem{Iyer:1994ys}
V.~Iyer and R.~M. Wald, ``Some properties of {N}oether charge and a proposal
  for dynamical black hole entropy,'' {\em Phys. Rev.} {\bf D50} (1994)
  846--864,
\href{http://www.arXiv.org/abs/gr-qc/9403028}{{\tt gr-qc/9403028}}.

\bibitem{Barnich:2006avcorr}
G.~Barnich and G.~Comp{\`e}re, ``Classical central extension for asymptotic
  symmetries at null infinity in three spacetime dimensions,'' {\em Class.
  Quant. Grav.} {\bf 24} (2007) F15,
  \href{http://www.arXiv.org/abs/gr-qc/0610130}{{\tt gr-qc/0610130}}.
Corrigendum: ibid 24 (2007) 3139.

\bibitem{Bergshoeff:2009hq}
E.~A. Bergshoeff, O.~Hohm, and P.~K. Townsend, ``{Massive Gravity in Three
  Dimensions},'' {\em Phys. Rev. Lett.} {\bf 102} (2009) 201301,
\href{http://www.arXiv.org/abs/0901.1766}{{\tt 0901.1766}}.

\bibitem{Bagchi:2009my}
A.~Bagchi and R.~Gopakumar, ``{Galilean Conformal Algebras and AdS/CFT},'' {\em
  JHEP} {\bf 07} (2009) 037,
\href{http://www.arXiv.org/abs/0902.1385}{{\tt 0902.1385}}.

\bibitem{Barnich:2010xq}
G.~Barnich, ``{A note on gauge systems from the point of view of Lie
  algebroids},'' {\em AIP Conf. Proc.} {\bf 1307} (2010) 7--18,
\href{http://www.arXiv.org/abs/1010.0899}{{\tt 1010.0899}}.

\bibitem{barnich:2011ct}
G.~Barnich and C.~Troessaert, ``{Supertranslations call for superrotations},''
  {\em Proceedings of Science CNCFG} {\bf 010} (2010)
\href{http://www.arXiv.org/abs/1102.4632}{{\tt 1102.4632}}.

\bibitem{Galperin:1985bj}
A.~Galperin, E.~Ivanov, V.~Ogievetsky, and E.~Sokatchev, ``{Harmonic
  Supergraphs. Green Functions},'' {\em Class.Quant.Grav.} {\bf 2} (1985) 601.

\bibitem{A.S.Galperin13674}
A.~S. Galperin, {\em Harmonic Superspace}.
\newblock New York : Cambridge University Press, 2001., Apr, 2011.

\bibitem{saidi:1990xx}
E.~H. Saidi and M.~Zakkari, ``{Harmonic distributions, Diff(S2), and Virasoro
  algebra},'' Tech. Rep. IC/90/257, ICTP, 1990.

\bibitem{PhysRevD.46.777}
E.~H. Saidi and M.~Zakkari, ``Virasoro algebra from harmonic superspace,'' {\em
  Phys. Rev. D} {\bf 46} (Jul, 1992) 777--785.

\bibitem{Geroch:1977aa}
R.~Geroch, ``Asymptotic structure of space-time,'' in {\em Symposium on the
  asymptotic structure of space-time}, P.~Esposito and L.~Witten, eds.,
  pp.~1--105.
\newblock Plenum, New York, 1977.

\bibitem{Ashtekar:1987tt}
A.~Ashtekar, ``{Asymptotic Quantization: Based on 1984 Naples Lectures},''.
  Naples, Italy: Bibliopolis (1987) 107 p. (Monographs and textbooks in
  physical science, 2).

\bibitem{PhysRevLett.10.66}
R.~Penrose, ``Asymptotic properties of fields and space-times,'' {\em Phys.
  Rev. Lett.} {\bf 10} (1963), no.~2, 66--68.

\end{thebibliography}

\def\cprime{$'$}
\providecommand{\href}[2]{#2}\begingroup\raggedright\endgroup

\end{document}